\documentclass[prd,aps,showpacs,preprintnumbers,amssymb]{revtex4}
\usepackage{axodraw}
\usepackage{color}
\usepackage{epsf}

\def\e3p{$\eta \rightarrow 3 \pi$}

\begin{document}
\title{%
\hfill{\normalsize\vbox{%
\hbox{}
 }}\\
{A theoretical estimate of the Higgs boson mass}}
\author{Renata Jora $^{\it \bf a}$~\footnote[1]{Email:
 rjora@theory.nipne.ro}}
 \affiliation{$^ {\bf \it a}$ National Institute of Physics and Nuclear Engineering, PO Box MG-6, Bucharest-Magurele, Romania.}
\date{\today}

\begin{abstract}
Using  the invariance of the electroweak vacuum in the presence of a background Z field we estimate the mass of the Higgs boson in the standard model at $m_H$=125.9 GeV.
\end{abstract}
\maketitle

\section{Introduction}
In an historical event the Atlas and CMS experiments at CERN recently announced the discovery of a Higgs like particle in the mass region 125 GeV \cite{CMS} or 126 GeV \cite{Atlas}.
 It is well known that the standard model \cite{Weinberg} does not predict the Higgs couplings or the mass so it is hard to have direct comparison between a theoretical estimate and the experiment.
  The usual treatment of theories with spontaneous symmetry breaking relies on the expansion of the lagrangian around the classical constant field which correspond to the minimum
   of the potential. Gauge bosons acquire masses through the Goldstone mechanism for which also the role of scalars is crucial. Despite the huge amount of information that one
    can get about the standard model from electroweak precision data the Higgs boson mass in the standard model remains up to today undetermined in a simple theoretical set-up.
    It is the lore of the present work to fill this gap.
We will illustrate our approach for the abelian Higgs model in section II. In section III we will give a detailed estimate of the mass of the Higgs boson in the standard model.
Section IV is dedicated to conclusions.

\section{The abelian Higgs model}

We start with the simplest model for spontaneous symmetry breaking \cite{Higgs}: a U(1) gauge theory coupled with a complex scalar field. The corresponding Lagrangian is:
\begin{eqnarray}
{\cal L}=-\frac{1}{4}(F_{\mu\nu})^2+|D_{\mu}\Phi|^2-V(\Phi),
\label{lagr45}
\end{eqnarray}

where,

\begin{eqnarray}
V(\Phi)=-\mu^2(\Phi^{\dagger}\Phi)+\lambda(\Phi^{\dagger}\Phi)^2.
\label{oloop8787}
\end{eqnarray}
We write the complex scalar field as,
\begin{eqnarray}
\Phi=\frac{1}{\sqrt{2}}(\Phi_0+\Phi_1+i\Phi_2).
\label{noty6768}
\end{eqnarray}

Here $\Phi_0$ is the vacuum expectation value of the scalar field $\Phi_0=\frac{\mu^2}{\lambda}$.
It is useful to write down all the terms that appear in the lagrangian with spontaneous symmetry breaking:

\begin{eqnarray}
&&|D_{\mu}\Phi|^2\rightarrow
\frac{1}{2}(\partial_{\mu}\Phi_1)^2+\frac{1}{2}(\partial_{\mu}\Phi_2)^2+
\frac{1}{2}g^2A_{\mu}A^{\mu}(\Phi_1^2+\Phi_2^2)
+
\nonumber\\
&&+gA^{\mu}\Phi_2\partial_{\mu}\Phi_1-gA^{\mu}\Phi_1\partial_{\mu}\Phi_2
+\frac{1}{2}g^2A_{\mu}A^{\mu}(\Phi_0^2+2\Phi_0\Phi_1)-
g A^{\mu}\Phi_0\partial_{\mu}\Phi_2
\label{usef3435}
\end{eqnarray}

In the Higgs potential the partition is as follows:

\begin{eqnarray}
&&-\frac{\mu^2}{2}\Phi_1^2-\frac{\mu^2}{2}\Phi_2^2+
\frac{\lambda}{4}\Phi_1^4+\frac{\lambda}{4}\Phi_2^4+\frac{\lambda}{2}\Phi_1^2\Phi_2^2+
\nonumber\\
&&-\frac{1}{2}\mu^2\Phi_0^2-\mu^2\Phi_1\Phi_0
+\frac{\lambda}{4}(\Phi_0^4+4\Phi_0^3\Phi_1+4\Phi_0\Phi_1^3+6\Phi_0^2\Phi_1^2)
+\frac{\lambda}{2}\Phi_2^2(\Phi_0^2+2\Phi_0\Phi_1)
\label{somres43534}
\end{eqnarray}

From these one can compute the effective action which identifies with the effective potential.

Assume now that we have spontaneous symmetry breaking and we decompose the gauge field into a background gauge field $B_{\mu}$ and a quantum gauge field $A_{\mu}$.
The background gauge field can be considered as slowly varying or approximately constant.

All terms previously listed will remain in the Lagrangian with the remark that the gauge field in Eq(\ref{usef3435}) should be identified with the quantum gauge field.
 Then the extra terms corresponding to the background gauge field are:
\begin{eqnarray}
&&\frac{1}{2}g^2(B_{\mu}B^{\mu}+2B_{\mu}A^{\mu})(\Phi_1^2+\Phi_2^2)
+g^2 B_{\mu}\Phi_2\partial^{\mu}\Phi_1 -B^{\nu}\Phi_1\partial_{\nu}\Phi_2+
\nonumber\\
&&\frac{1}{2}g^2B_{\mu}B^{\mu}(\Phi_0^2+2\Phi_1\Phi_0)
+g^2B_{\mu}A^{\mu}(\Phi_0^2+2\Phi_1\Phi_0)-g B_{\mu}\Phi_0\partial^{\mu}\Phi_2.
\label{res55546}
\end{eqnarray}

From Eq (\ref{res55546}) we can select the contribution,
\begin{eqnarray}
\frac{1}{2}g^2B_{\mu}B^{\mu}\Phi_0^2+ g^2B_{\mu}A^{\mu}\Phi_0^2
\label{tr5646}
\end{eqnarray}
which will lead at tree level to,
\vspace{0.5cm}
\begin{center}
\begin{picture}(200,50)(0,0)
\Photon(10,30)(70,30){3}{4}
\Vertex(40,30){2}
\Text (80,30)[]{+}
\Photon(90,30)(180,30){3}{12}
\Vertex(120,30){2}
\Vertex(150,30){2}
\Text(190,30)[]{=}
\Text(90,0)[]{$ig^2\Phi_0^2g^{\mu\nu}+ (i)g^2\Phi_0^2\frac{-i}{-g^2\Phi_0^2}(i)g^2\Phi_0^2g^{\mu\nu}=0$}
\end{picture}
\vspace{0.5cm}
\\{\sl Fig.1. The cancelation of terms.}
\end{center}

and thus to the cancelation of these terms.

In rest we need to determine only those terms which contain the background gauge field and the Higgs field:
\begin{eqnarray}
2g^2B_{\mu}A^{\mu}\Phi_1\Phi_0+g^2B_{\mu}B^{\mu}\Phi_1^2
\label{ter445367}
\end{eqnarray}

The second term in Eq(\ref{ter445367}) can be regarded as an extra mass term for the Higgs boson.
Besides  this at one loop the only term that contributes is the first term in Eq(\ref{ter445367}).
\vspace{0.5cm}
\begin{center}
\begin{picture}(200,100)(0,0)
\Photon(10,50)(110,50){3}{5}
\Vertex(40,50){1}
\Vertex(80,50){1}
\Text(20,60)[]{$B_{\mu}$}
\Text(100,60)[]{$B_{\mu}$}
\CArc(60,50)(20,0,180)
\end{picture}
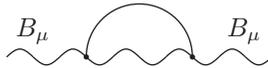
\\{\sl Fig.2. One diagram that contributes to the effective action.}
\end{center}

The main point is the determine the one loop effective action for a theory with spontaneous symmetry breaking in the presence of a background gauge field.
 For the simple case at hand we expect that this action will contain the result of the diagram in Fig.2 plus the usual one loop effective
 scalar potential where the masses of the scalars will have an extra  contribution proportional to
$B_{\mu}B^{\mu}g^2$.

Assume that Z[J,K] is the generating functional for a spontaneously broken gauge theories where J, K are generic sources corresponding to the scalars and gauge fields respectively.
Now we introduce a background gauge field $B_{\mu}$. The new generating functional reads \cite{Abbott}:
\begin{eqnarray}
&&Z'[J,K,B]=\int \delta Q \delta A \exp[i[S[Q,A+B]+JQ+KA]]
=
\nonumber\\
&&\int \delta Q \delta A\exp[iS[Q,A]+iJQ+iK(A-B)]=
Z[J,K]\exp[-iKB]
\label{someus54646}
\end{eqnarray}

The vacuum expectation value of the scalar field before introducing the background gauge field is given by the expression:
\begin{eqnarray}
Q_J=\frac{-i\ln Z[J,K]}{\delta J}
\label{r4546}
\end{eqnarray}
 whereas in the presence of the background gauge field is
 \begin{eqnarray}
 Q_J'=\frac{-i\ln Z[J,K]}{\delta J}-\frac{\delta K B}{\delta J}=\frac{-i\ln Z[J,K]}{\delta J}.
 \label{res435242}
 \end{eqnarray}

 Thus  Eq(\ref{res435242}) states that the vacuum expectation of the scalar field is preserved also in the presence of the background gauge field (Note that this is true even if
  the gauge source depends on the minimum of the scalar field).
  By setting the sources to zero one obtains
the minimum of the effective action (or potential):
\begin{eqnarray}
\frac{\delta \Gamma[B,\Phi_{cl}]}{\delta \Phi_{cl}}=
\frac{\delta \Gamma[\Phi_{cl}]}{\delta \Phi_{cl}}
\label{ewm777}
\end{eqnarray}

Here we denote by $\Gamma[B,\Phi_{cl}]$ the effective action in the presence of the background gauge field whereas $\Gamma[\Phi_{cl}]$ is the effective action in the absence of it.
As the background gauge field is arbitrary one should expand $\Gamma[B,\Phi_{cl}]$ in terms of B. In what follows we will consider only terms of order $B^2$ and neglect the higher order
contributions.
Applied to our abelian Higgs model this means that the sum of the term in Fig.2 plus the term in the scalar potential proportional to $B_{\mu}B^{\mu}$ will give zero contribution
to the vacuum  expectation of the scalar. From this one can derive various constraints and even estimate the Higgs boson mass.
We will illustrate in detail how this works for the standard model in the next section.

\section{ Higgs boson mass in the standard model}

We will work with a Higgs potential of the form:
\begin{eqnarray}
V=-m^2\Phi^{\dagger}\Phi+\lambda(\Phi^{\dagger}\Phi)^2
\label{Higgs54646}
\end{eqnarray}

which displays spontaneous symmetry breaking as the scalar develops the vacuum expectation value:
\begin{eqnarray}
\langle 0 |\Phi| 0\rangle=
\left[
\begin{array}{cc}
0\\
\frac{\Phi_0}{\sqrt{2}}
\end{array}
\right]
\label{veve435}
\end{eqnarray}.

The next step is to introduce a background gauge field for the Z boson, for the photon or the $W^{\pm}$ bosons.  The case for the W bosons is more intricate and we will not discuss it here. Assume that we introduce only a background gauge field for the photon. This does not couple to the Higgs so it cannot help us with determining the mass of the Higgs boson. However it has almost symmetric interactions (with different couplings) with the W bosons as the Z boson as it is evident from the interaction term in the Lagrangian:
\begin{eqnarray}
-D^{\dagger\mu}W^{-\nu}D_{\mu}W_{\nu}^{+}+D^{\dagger\mu}W^{-\nu}D_{\nu}W_{\mu}^{+}
\label{lagr65757}
\end{eqnarray}
where,
\begin{eqnarray}
D_{\mu}=\partial_{\mu}-ie(A_{\mu}+\cot{\theta_W}Z_{\mu}).
\label{terr54646}
\end{eqnarray}
The contribution of the background photon field to the effective action will thus contain only the interactions depicted above and the interaction with the fermions.
This means that the W bosons will also give per total zero contribution to the vacuum expectation value of the scalars
(through the derivative of the corresponding term in the effective action). By the symmetry of the W-photon and W-Z interactions we conclude that
if one introduces a background gauge field only for the Z boson those terms that contain Z interaction to the W should not be taken into consideration.

Then the effective action will be simply the sum of the regular effective potential \cite{Sher},\cite{Jones},
\begin{eqnarray}
V(\Phi)&&=-\frac{1}{2}m^2\Phi^2+\frac{\lambda}{4}\Phi^4+\frac{1}{64\pi^2}(-m^2-g^2B^2+3\lambda\Phi^2)^2[\ln(\frac{-m^2-g^2B^2+3\lambda\Phi^22}{\mu_0^2})-\frac{3}{2}]+
\nonumber\\
&&+\frac{3}{64\pi^2}(-m^2-g^2B^2+\lambda\Phi^2)^2[\ln(\frac{-m^2-g^2B^2+\lambda\Phi^2}{\mu_0^2})-\frac{3}{2}]+
\frac{3}{64\pi^2}(\frac{1}{4}(g^2+g'^2)\Phi^2)^2[\ln(\frac{\frac{1}{4}(g^2+g'^2)\Phi^2}{\mu_0^2})-\frac{5}{6}]+
\nonumber\\
&&+\frac{3}{32\pi^2}(\frac{1}{4}g^4\Phi^2)[\ln(\frac{\frac{1}{4}g^2\Phi^2}{\mu_0^2})-\frac{5}{6}]-
\frac{3}{16\pi^2}(\frac{1}{2}g_t^2\Phi^2)^2[\ln\frac{\frac{1}{2}g_t^2\Phi^2}{\mu_0^2}-\frac{3}{2}]
\label{oneloop9898}
\end{eqnarray}

and the result of the one loop diagram (see Fig.2) corresponding to the vertex $2B_{\mu}Z^{\mu}\Phi_0\Phi_1$ where $\Phi_1$ is the Higgs boson,
\begin{eqnarray}
&&\Gamma_1[B^2]=\frac{6B^2}{16\pi^2}g_z^2\Phi^2(1+\frac{1}{-m^2-g_z^2\Phi^2-g_z^2B^2}[g_z^2\Phi^2\ln(\frac{gz^2\Phi^2}{\mu_0^2})-
\nonumber\\
&&-(-m^2+3\lambda\Phi^2-g_z^2B^2)\ln(\frac{-m^2+3\lambda\Phi^2-g_z^2B^2}{\mu_0^2})].
\label{loop87878}
\end{eqnarray}

Here $g_z^2=\frac{g^2+g'^2}{4}$.
There is also the one loop fermion contribution (we consider only the top quark):
\begin{eqnarray}
\Gamma_2[B^2]=B^2[6\frac{g^2g_t^2}{8\cos^2\theta_W}[(1-(1-\frac{8}{3}\sin^2\theta_W)^2)+2\ln[\frac{g_t^2\Phi^2}{2\mu_0^2}]]
\label{quarkt}
\end{eqnarray}

The full one loop effective potential is thus:
\begin{eqnarray}
\Gamma[B,\Phi]=\Gamma_1[B^2]+\Gamma_2[B^2]-V_{eff}[\Phi,B]+ {\rm higher\,order\,terms\,in\,B}.
\label{efe54646}
\end{eqnarray}

We denote $\Gamma[B^2]=\Gamma_1[B^2]+\Gamma_2[B^2]$.
The invariance of the vacuum in the presence of background gauge field translates into the constraint:
\begin{eqnarray}
\frac{d\Gamma[B,\Phi]}{d\Phi d B^2}|_{B^2=0}=0.
\label{43535}
\end{eqnarray}
where we used the expansion of the effective action in powers of $B^2$
( The sum over the $B^{2n}$ terms can be obtained formally by integrating the $B^2$ term so the condition (\ref{43535}) suffices).

 In what follows we will apply the constraint in Eq(\ref{43535}).
Furthermore we will simply take $m_H^2=2\lambda \Phi_0^2$ which is true apart for some small corrections; $\mu_0^2=m_H^2$ and the electroweak values($g_z^2=\frac{M_Z^2}{\Phi_0^2}$,
 for the coupling constants with the notation: $g_z^2=\frac{M_Z^2}{\Phi_0^2}$, $g_w^2=\frac{M_W^2}{\Phi_0^2}$, $g_y^2=\frac{m_t^2}{\Phi_0^2}$). Then
 Eq(\ref{43535}) reduces to:

 \begin{eqnarray}
 &&-6g_z^2/(g_z^2-2\lambda)^2[-\lambda(g_z^2-2\lambda)+g_z^2(g_z^2-\lambda)\ln(\frac{g_z^2}{2\lambda})+
 \nonumber\\
&&\frac{12g_y^2g_w^2}{\cos^2\theta_W}\ln(\frac{g_y^2}{2\lambda})+\frac{6g_y^2g_w^2}{\cos^2\theta_W}[1-(1-\frac{8}{3}\sin^2\theta_W)^2)]=0
\label{res546}
\end{eqnarray}

This leads to the following result for the coupling and the mass of the Higgs boson: $\lambda=0.13$; $m_H=125.9$. In this estimate we also used a small correction coming form the bottom quark.

\section{Discussion}
 Our approach is based on the introduction of a background gauge field in a Higgs theory after the spontaneous symmetry breaking.
 The background gauge is considered slowly varying such that any derivative term can be neglected.
 We show that in particular the vacuum expectation value of the scalar field remains unchanged by this transformation.
 We then apply the constraint deduced from this invariance to the standard model. In this case the most advantageous choice is to introduce a background gauge field for the Z boson.
 We obtained a mass for the standard model Higgs boson of $m_H=125.9$ in excellent agreement with the experimental results.  Our result is
 susceptible to corrections coming from the running of the coupling constants and fields. We expect these corrections to be small.

\section*{Acknowledgments} \vskip -.5cm
I am happy to thank J. Schechter and Marc Sher for useful comments on the manuscript.
This work has been supported by PN 09370102/2009.

\end{document}